\documentclass[submission,copyright,creativecommons]{eptcs}
\providecommand{\event}{LFSA 2011}
\usepackage{breakurl}

\usepackage{eqalignno}
\usepackage{amsmath}
\usepackage{amssymb}

\title{A Logical Framework for Set Theories}

\author{Arnon Avron
\institute{School of Computer Science\\
Tel Aviv University, Tel Aviv 69978, Israel}
\email{aa@math.tau.ac.il}}

\def\titlerunning{A Logical Framework for Set Theories}
\def\authorrunning{A. Avron}
\begin{document}
\maketitle

\newtheorem{definition} {Definition}
 \newtheorem{theorem}{Theorem}
 \newtheorem{example}{Example}
 \newtheorem{corollary}{Corollary}
 \newtheorem{lemma}{Lemma}
 \newtheorem{proposition}{Proposition}
 \newtheorem{note}{Note}

\newcommand{\notation}[1]{\begin{Notation} {\rm #1} \end{Notation}}
\newcommand{\tup}[1]{\langle #1 \rangle}

\newcommand{\topic}[2]{\section{#1} \label{#2}}
\newcommand{\subtopic}[2]{\subsection{#1} \label{#2}}
\newcommand{\subsubtopic}[2]{\subsubsection{#1} \label{#2}}

\newcommand{\rhdx}{\triangleright}

\def\RR{\mathbb R}
\def\hatx{\widehat x}
\def\note{\smallskip\noindent{\bf Note.\enskip}}
\def\notes{\smallskip\noindent{\bf Notes.\enskip}}
\def\Del{{\Delta}}
\def\Gam{{\Gamma}}
\def\lam{{\lambda}}
\def\Ome{{\Omega}}
\def\alp{{\alpha}}
\def\sig{{\sigma}}
\def\arrowsim{\mathrel{\vcenter{
\offinterlineskip{\hbox{$\scriptstyle\sim$}
\hbox{$\scriptscriptstyle\rightarrow$}}}}}
\def\to{\rightarrow}
\def\lol{\mathrel{\vcenter{\offinterlineskip{\hbox{$<$}
\hbox{$<$}}}}}
\def\nek{,\ldots ,}
\def\pr{\smallskip\noindent{\bf Proof:\quad}}
\def\oS{{\overline S}}
\def\oA{{\overline A}}
\def\oF{{\overline F}}
\def\NN{{\mathbb N}}
\newcommand{\vd}{\vdash}
\newcommand{\de}{\Delta}
\newcommand{\w}{\wedge}
\newcommand{\du}{\underline{D}}
\newcommand{\sub}{\subseteq}
\newcommand{\fo}{\forall}
\newcommand{\g}{\Gamma}
\newcommand{\ld}{\ldots}
\newcommand{\ra}{\rightarrow}
\newcommand{\s}{\sequent}
\newcommand{\R}{\Rightarrow}
\newcommand{\lw}{\leftrightarrow}
\newcommand{\ex}{\exists}
\newcommand{\Di}{\Diamond}
\newcommand{\cre}[1]{\vd_{#1}}
\newcommand{\bi}{\begin{itemize}}
\newcommand{\ei}{\end{itemize}}
\newcommand{\be}{\begin{enumerate}}
\newcommand{\ee}{\end{enumerate}}
\newcommand{\bd}{\begin{description}}
\newcommand{\ed}{\end{description}}
\newcommand{\fe}{\varphi}
\newcommand{\bs}{\bigskip}
\newcommand{\ms}{\medskip}

\begin{abstract}
Axiomatic set theory is almost universally accepted as the basic
theory which provides the foundations of mathematics, and in which
the whole of present day mathematics can 
be developed. As such, it is the most natural framework for 
Mathematical Knowledge Management.
However, in order to be used for this task
it is necessary to overcome serious gaps that exist between the
``official'' formulations of set theory
(as given e.g.  by formal set theory $ZF$)
and actual mathematical practice.

In this work we present a new unified framework for 
formalizations of axiomatic set theories
of different strength, from rudimentary set theory to full $ZF$. It
allows the use of set terms, but provides a {\em static} check of 
their validity.  Like the inconsistent ``ideal calculus" for set theory,
it is essentially based on just two
set-theoretical principles: extensionality and
comprehension (to which we add $\in$-induction and optionally
the axiom of choice).  Comprehension is formulated as:
$x\in\{x\mid\fe\}\leftrightarrow\fe$, where $\{x\mid\fe\}$
is a legal set term of the theory. 
In order for $\{x\mid\fe\}$
to be legal, $\fe$ should be {\em safe} with respect to $\{x\}$,
where safety is a relation between formulas and finite sets of variables.
The various systems we consider differ from each other mainly with 
respect to the safety relations they employ. These relations are all
defined purely syntactically (using an induction on the logical structure
of formulas).  The basic one 
is based on the safety relation which implicitly underlies
commercial query languages for relational database systems (like SQL).

\end{abstract}

\section{Introduction}

Axiomatic set theory is almost universally accepted as the basic
theory which provides the foundations of mathematics, and in which
the whole of present day mathematics can (and  many say: should)
be developed. As such, it is the most natural framework for MKM
(Mathematical Knowledge Management).
Moreover: as is emphasized and demonstrated in
\cite{COP01}, set theory
has not only a great pragmatic advantage as a basic language for mathematical
discourse, but it
also has a great computational potential as a basis for
specification languages, declarative programming, and
proof verifiers. However, in order to be used for any of these tasks
it is necessary to overcome the following serious gaps that exist between the
``official'' formulations of set theory
(as given e.g.  by Zermelo Fr\"ankel Set Theory $ZF$; see e.g.~\cite{FBL73}).
and actual mathematical practice:

\bi
\item ZF treats all the mathematical objects on a par, and so hid
the computational significance of many of them. Thus although
certain functions are first-class citizens in many programming languages,
in set theory they are just ``infinite sets", and ZF
in its usual presentation is an extremely poor framework
for computing with such sets (or handling them in a constructive way).

\item Full ZF is far too strong for core mathematics, which practically
deals only with a small fraction of the set-theoretical ``universe". It
is obvious that much weaker systems, corresponding to universes which are
smaller, {\em more effective},
and better suited for computations , would do (presumably,
such weaker systems will also be  easier to mechanize).

\ei

The goal of this paper is to present a 
unified, user-friendly framework 
(originally developed in \cite{Av07})
for formalizations of axiomatic set theories
of different strength, from rudimentary set theory
to full ZF. 
Our framework makes it  possible to employ in a natural way
all the usual set notations and constructs
as found in textbooks on naive or axiomatic set theory (and
{\em only} such notations). Another important feature of
this framework is that its set of closed terms suffices
for denoting every concrete set (including infinite ones!)
that might be needed in applications, as well as for
{\em computations} with sets.

\ms

Perhaps the most important  problem which is solved in our framework is that
official formalizations of axiomatic set theories
in almost all textbooks are based on some {\em standard} first-order languages.
In such languages terms are variables, constants, and sometimes
function applications (like $x\cap y$). What is {\em not}
available in the {\em official} languages of these formalizations
is the use of set terms of the form ($\{x\mid\fe\}$).
As a result, already the formulation of the axioms is quite cumbersome,
and even the formalization of elementary proofs becomes
something practically incomprehensible. In contrast,
{\em all} modern texts in all areas of mathematics
(including set theory itself)
use such terms extensively. For the purpose of mechanizing
real mathematical practice and for automated or
interactive theorem proving,  it is therefore important to have
formalizations of ZF and related systems which allow the use of such terms.
Now, set terms {\em are} used  in all textbooks
on first-order set theories, as well as in several computerized systems.
However, whenever they are intended to denote {\em sets} (rather than classes)
 they are introduced (at least partially) in a
{\em dynamic} way, based  for example on the ``extension by definitions"
procedure (see \cite{Sh67}, Sect. 4.6): In order to be able to introduce some
set term for a {\em set} (as well as a new operation on {\em sets})
it is  necessary first to justify
this introduction by {\em proving} a corresponding existence
{\em theorem}.
The very useful complete separation
we have in first-order logic between
the (easy) check whether a given expression is a  well-formed
term or formula, and  the (difficult) check whether it is
a theorem, is thus lost. By analogy to programs: texts 
in such  dynamic languages can
only be ``interpreted", but  not ``compiled".
In contrast, a  crucial feature of our framework is 
that although it makes extensive use of set terms, 
the languages used in it are all {\em static}: the task
of verifying that a given term or  formula is well-formed
is decidable, easily mechanizable, and completely separated from any task
connected with proving theorems (like finding proofs or checking
validity of given ones). Expanding the language is allowed
only through {\em explicit} definitions (i.e. new valid expressions
of an extended language will just be abbreviations for expressions
in the original language). This feature has the same obvious
advantages that static type-checking has over
dynamic type-checking.

\medskip

Two other important features of the framework we propose are:

\bi
\item It provides a unified treatment of two important
subjects of set theory: axiomatization and absoluteness (the latter
is a crucial issue in independence proofs and in the study of models
of set theories -- see e.g. \cite{Ku80}).
In the usual approaches these subjects are
completely separated. Absoluteness is investigated mainly from a
syntactic point of view, axiomatizations -- from a semantic one.
Here both are given the same syntactic treatment.
In fact, the basis of the framework is its formulation of 
rudimentary set theory,
in which only terms for absolute sets are allowed.
The other set theories  are obtained from it by
small changes in the syntactic definitions. 
\item Most of our systems (including the one which is
equivalent to $ZF$) have the remarkable property that every set or function
that is implicitly definable in them  already has a term
in the corresponding language which denotes it.
More precisely: if $\fe(x,y_1,\ld,y_n)$
is a formula such that $\fo y_1,\ld,y_n\exists ! x\fe$ is provable,
then there is a term $t(y_1,\ld,y_n)$ such that
$\fe(y_1,\ld,y_n,t(y_1,\ld,y_n))$ is provable. Hence, there is no need
at all for the procedure of extension by definitions  (and introduction
of new symbols is completely reduced
to using {\em abbreviations}).
\ei

\section{The Major Ideas}

Our basic assumption  is that the  sets which are interesting
from a computational point of view are those which can be
{\em defined}  in the form $\{x\mid\fe\}$ 
using a formula  $\fe$ in some, intuitively
meaningful, formal language. Of course, the paradoxes of naive
set theory have shown that
not every formula of such a language can be used  for defining sets.
Accordingly, the crucial
question is: what formulas are ``safe"
for this task, and more generally:
what formulas can be taken as defining a {\em construction}
of a set from given objects (including other sets)?
Various set theories provide different answers to this question.
These answers are usually
guided by semantic intuitions (like the limitation of size doctrine
\cite{FBL73}). 
Since here we aim at
a computerized system, we shall translate the
various semantic principles into  {\em syntactic} (and in our opinion,
less ad-hoc) constraints on the logical form of formulas.
For this,  we combine ideas from three seemingly very different
sources:

\bd
\item[Set Theory] G\"odel's classical work \cite{Go40}
on the constructible universe $L$
is best known for its use in
consistency and independence proofs. However, it is of course
of great interest also for the study of the general
notion of constructions with sets.
Thus for characterizing the ``constructible sets" G\"odel identified a
set of operations on sets (which we may call ``computable"), that
can be used for ``effectively" constructing new sets
from given ones.
For example, binary union and intersection
are ``effective", while the powerset operation is not.
G\"odel has provided
a finite list of basic operations,
from which all other ``effective" (for his purposes)
constructions can be obtained
through compositions.
Another very important idea
which was introduced in \cite{Go40} is {\em absoluteness} --- a key
property (see \cite{Ku80})
of formulas which are used for defining ``constructible sets".
Roughly, a formula 
is absolute if its truth value in a transitive class
$M$, for some assignment $v$ of objects from $M$
to its free variables, depends only on $v$, but not on $M$.
\item[Formal arithmetic]
Absoluteness is not a decidable property. Therefore a certain set $\Delta_0$ of
absolute formulas
is extensively used in set theory as a syntactically defined approximation.
Now a similar  set $\Delta_0$ of  formulas (also called in \cite{Sm92}
``bounded formulas" or ``$\Sigma_0$-formulas")
which has {\em exactly the same definition} (except that $\in$ is replaced
by $<$) is used in formal arithmetic in order to characterize the
{\em decidable} and the semi-decidable (r.e.) relations on the
natural numbers. This  fact hints at  an intimate
connection (investigated in \cite{Av08}) between absoluteness/constructibility
and decidability/computability. 

\item[Relational database theory:] The importance of computations
with sets to this area is obvious: to provide
an answer to a query in a relational database, a computation
should be made in which  the input is a finite set
of finite sets of tuples (the ``tables" of the database), and the output
should also be a finite
set of tuples. In other words: the computation is done with
(finite) sets. 
Accordingly, for {\em effective} computations
with finite relations some finite
set of basic operations has been identified in database theory,
and this basic set defines (via composition) what is called there
``the relational algebra" (\cite{AHV95,Ul88}).
Interestingly,
there is a lot of similarity between
the list of operations used in the  relational algebra
and G\"odel's list of basic operations mentioned above.
However, much more important is again the strong connection
(observed in (\cite{Av04,Av08}) between
the notion of absoluteness used in set theory, and
the notion of {\em domain independence} (\cite{AHV95,Ul88})
used in database theory,
and practically serving
as its  counterpart of the notion of computability.
A query in a database can be construe as a formula $\fe$
in the language of set theory, augmented with constants for the
relations in the database. The answer to such query is the set of all
$n$-tuples that satisfy $\fe$, given the interpretations
provided by the database for the extra constants (here $n$ is the number
of free variables in $\fe$. 
If $n=0$ then the answer to the query is either ``yes" or ``no").
A domain-independent (d.i.) query is a query the answer to which depends only
on the information
included in the database, and on the objects which are mentioned
in the query. Only such queries are considered meaningful. Moreover:
the answer to such queries is always finite and {\em computable}.
Therefore practical database query languages (like SQL)
are designed so that only d.i. queries can be formulated
in them, and each such query
language is based on some syntactic criteria that ensure this property.
In order to give these criteria a concise logical characterization,
and in order to unify the notions of absoluteness
and domain-independence,  the formula {\em property} of d.i.
was turned in \cite{Av04,Av08}
into a {\em safety relation} $\succ$ between a formula $\fe$
and finite subsets of
$Fv(\fe)$. The intuitive meaning of
``$\fe(x_1\nek x_n,y_1\nek y_k)\succ\{x_1\nek x_n\}$" in databases is:
``$\fe(x_1\nek x_n,d_1\nek d_k)$ is d.i. for all values $d_1\nek d_k$".
In particular, $\fe\succ\emptyset$ if $\fe$ is {\em absolute} in the sense
of axiomatic set theory.

\ed

In view of the connections 
between ``absolute" and ``decidable" and between 
``domain-independent"
and ``computable",  (or ``constructible"), 
in the realm of sets we shall  intuitively take
the meaning of ``$\fe(x_1\nek x_n,y_1\nek y_k)\succ\{x_1\nek x_n\}$" to be:
``The collection $\{\tup{x_1\nek x_n}\mid\fe\}$ is an acceptable
set for all acceptable values of $y_1\nek y_k$, and it can
be {\em constructed} from these values".  The differences between
the strength of systems is intuitively due to different
interpretations of the vague notions of
``acceptable" and ``can be constructed". At least
in the  basic systems, but also in
some of the less basic ones, a crucial part of the meaning of both concepts
is the  demand that $\{\tup{x_1\nek x_n}\mid\fe\}$ 
is ``domain independent" in a sense close to that used in database theory,
i.e.: that $\fe$ determines this collection  in an absolute way, independent
of the extension of the ``surrounding universe" $V$. In particular:
$\fe\succ\emptyset$ implies in such set theories that $\fe$ is absolute
(in the set-theoretical sense mentioned above).

\section{A Description of the General Framework}

\subsection{Languages}
\label{Languages}

In our framework a language $L$ for a set theory $S$
should be based on some first-order signature
$\sigma$ which includes the binary predicate
symbols $\in$ and $=$. 
Moreover: it should be 
introduced using a simultaneous recursive definition of 
the following three components: 
its set of terms, its set of formulas, 
and the {\em safety relation} $\succ$  that it  uses between
formulas and finite sets of variables.
The recursive definition of  these components  includes
at least the following conditions: 

\begin{description}
\item[Terms:] \
\bi
\item Every variable and every constant of $\sigma$ is a term.
\item If $f$ is an $n$-ary function symbol of $\sigma$,
and $t_1,\ldots,t_n$ are terms, then $f(t_1,\ldots,t_n)$ is a term.
\item If $x$ is a variable, and $\fe$ is a formula such that
$\fe\succ\{x\}$,  then $\{x\mid\fe\}$ is a term.
\ei
\item[Formulas:] \
\bi
\item If $P$ is an $n$-ary predicate symbol of $\sigma$,
and $t_1,\ldots,t_n$ are terms, then $P(t_1,\ldots,t_n)$ is an
atomic formula.
\item If $\fe$ and $\psi$ are formulas, and $x$ is a variable, then
$\neg\fe$, $(\fe\w \psi)$, $(\fe\vee\psi)$, 
and $\exists x\fe$ are formulas. In an intuitionistic system
so are also $(\fe\to \psi)$ and $\forall x\fe$ (but in the
classical case $\to$  and $\forall$
are better taken as defined in terms of $\neg$, $\w$, and $\exists$).
\item An optional construct which may be useful in our
framework and is not available in first-order languages
is the transitive closure operation $TC$. If it 
is included, then $(TC_{x,y}\varphi)(t,s)$ is a formula whenever
$\varphi$ is a formula, $x,y$ are distinct variables, and $t,s$ are terms.
In this  formula all occurrences of $x$ and $y$ in $\varphi$ are bound.
The intended meaning of  $(TC_{x,y}\fe)(t,s)$ is
the ``disjunction":
$\fe\{s/x,t/y\}\vee\exists w_1(\fe\{s/x,w_1/y\})\wedge \fe\{w_1/x,t/y\})\vee
\exists w_1\exists w_2(\fe\{s/x,w_1/y\}\wedge \fe\{w_1/x,w_2/y\}
\wedge \fe\{w_2/x,t/y\})\vee\ldots$
(where $w_1,w_2\nek$ are all new variables)). 
\ei

\item[Safety Relation:] \
\bi
\item $\fe\succ_{}\emptyset$ if $\fe$ is atomic.
\item $\fe\succ_{}\{x\}$ if $\fe\in\{x=t,t=x,x\in x,x\in t\}$, and
 $x\not\in Fv(t)$.
\item $\neg\fe\succ_{}\emptyset$ if $\fe\succ_{}\emptyset$.
\item $\fe\vee \psi\succ_{}X$
if $\fe\succ_{}X$ and $\psi\succ_{}X$.
\item $\fe\wedge \psi\succ_{}X\cup Y$ if
$\fe\succ_{}X$, $\psi\succ_{}Y$ and 
$Y\cap Fv(\fe)=\emptyset$,  or $X\cap Fv(\psi)=\emptyset$.
\item $\exists y \fe\succ_{}X-\{y\}$ if $y\in X$ and $\fe\succ_{}X$.
\item $\forall x(\fe\to\psi)\succ\emptyset$ if $\fe\succ\{x\}$
and $\psi\succ\emptyset$\footnote{In the classical case this condition
is derivable from the others.}
\item If  $TC$ is included in the language then
$(TC_{x,y}\fe)(x,y)\succ X$
if $\fe\succ X$, and $\{x,y\}\cap X\neq\emptyset$.
\ei
\ed

\noindent
{\bf Notes:}
\be
\item The clauses concerning $\succ$ form a generalization
(and simplification) of the definition of ``syntactically safe"
formulas from \cite{Ul88} (see \cite{Av04,Av08,Av10}). The passage
from the property of domain independence to the safety relation
is mainly needed for  an appropriate handling of conjunction.
\item Recalling the intended intuitive meaning(s) of our safety relations,
is not difficult to see that any safety relation $\succ$
should satisfy the conditions listed above. As an example, we explain
the most complicated of them: the one connected with $\w$. Assume
for simplicity that $\theta=\fe\w\psi$, where $Fv(\fe)=\{x,z\}, 
Fv(\psi)=\{x,y,z\},
\fe\succ\{x\}$, and $\psi\succ\{y\}$. Given some ``acceptable" 
set $c$, we should show that the 
collection $E(c)$ of all $\tup{x,y}$ such that $\theta(x,y,c)$ should
also be taken as ``acceptable". Now the assumption that $\fe\succ\{x\}$
implies that the collection $Z(c)$ of all $x$ such
that $\fe(x,c)$ is ``acceptable". 
In turn, the the assumption that $\psi\succ\{y\}$
implies that for every $d$ in this set, the collection $W(c,d)$ of all $y$
such that $\psi(d,y,c)$ is  ``acceptable" . Since $E(c)$ is the union for
$d\in Z(c)$ of the sets $\{d\}\times W(c,d)$, it is 
constructible from ``acceptable" sets using G\"odel's basic operations
mentioned above, and so it too should intuitively be ``acceptable"
in any reasonable set theory. What is more, if $Z(c)$ is
``constructible" from $c$ (in an absolute way),
and $W(c,d)$ is ``constructible" from $c$ and $d$ 
(in an absolute way), then
this argument shows that $E(c)$ 
is ``constructible" from $c$ (in an absolute way) as well.
\item The recursive definition of $\succ$ 
should ensure that $\succ$ has the following properties:
\bi
\item If $\fe\succ X$ then  $X\subseteq Fv(\fe)$.
\item If $\fe\succ X$ and $Z\subseteq X$, then $\fe\succ Z$.
\item If $\fe\succ \{x_1,\ldots,x_n\}$, $v_1,\ldots v_n$ are
$n$ distinct variables not occurring in $\fe$,  and
$\fe^\prime$ is obtained from $\fe$ by replacing all 
occurrences of $x_i$ by $v_i$ ($i=1\nek n$), then
$\fe^\prime\succ \{v_1,\ldots,v_n\}$
\ei
It is easy to verify that all the safety relations used in the
examples below have these properties, and
so there is no need to add corresponding clauses
to their definitions (but this might not be
the case in general).
\ee

\subsection{Logics}

Our framework allows the use of any logic
that is based one of the two languages it employs
(with classical and intuitionistic logics as the natural choices).
One should note however the following points:

\be
\item Our languages provide  much richer classes of terms than those
allowed in orthodox first-order systems. In particular: a variable
can be bound in them within a term. The notion of a term being
free for substitution is generalized accordingly (also
for substitutions within terms!). As usual this amounts to avoiding the
capture of free variables within the scope of an operator
which binds them. Otherwise the rules/axioms concerning the
quantifiers and terms remain unchanged (for example:
$\fe[x\mapsto t]\to\exists x\fe$ is valid for 
{\em every} term $t$ which is free
for $x$ in $\fe$). 

\item The rule of  $\alpha$-conversion (change of bound variables)
should be available in 
the logic.

\item The substitution of equals for equals should be allowed within any
context (under the usual conditions concerning bound variables).
The same should apply for the substitution
of a formula for an equivalent formula in any context in
which the substitution makes sense. In particular,
the following schema should be  valid whenever 
$\{x\mid\fe\}$ and $\{x\mid\psi\}$
are legal terms:
$$\fo x(\fe\leftrightarrow\psi)\to\{x\mid\fe\}=\{x\mid\psi\}$$

\item The set of valid formulas of first-order languages
enriched with the TC operator
is not even arithmetical. Hence no sound and complete formal
system for it is possible. It follows that
only  appropriate formal approximations of the intended
underlying  logic may be used in practice. The best
known  approximation is the one given in \cite{My52},
using a Hilbert-type system. An equivalent 
Gentzen-type formulation ({\em with} cuts) 
has been given in \cite{Av03}. In that system mathematical induction
is presented as the following  logical rule:

\[\frac{\Gam, \psi, \varphi
\Rightarrow \Del, \psi[x\mapsto y]}{\Gam,
\psi[x\mapsto s], (TC_{x,y}\varphi) (s,t)
\Rightarrow \Del, \psi[x\mapsto t]}\]
where $x$ and $y$ are not
free in $\Gam, \Del$, and
$y$ is not free in  $\psi$.

\ee

\subsection{Axioms}
\label{Axioms}

The main part of all systems in our framework consists of the following
axioms and axiom schemes (our version of the
``ideal calculus" \cite{FBL73},
 augmented with the assumption that we are
dealing with the cumulative universe):
\bs
\bd
\item[Extensionality:]\
\ms
\bi
\item $\fo y (y=\{x\mid x\in y\})$
\ei
\ms
\item[Comprehension Schema:] \
\ms
\bi
\item $\forall x(x\in\{x\mid\fe\}\leftrightarrow\fe)$
\ms
\ei
\item[The Regularity Schema ($\in$-induction):] \
\ms
\bi
\item $(\forall x(\forall y(y\in x\to \fe[x\mapsto y])\to \fe))\to \forall x\fe$
\ms
\ei
\ed
\noindent
{\bf Notes:}
\be
\item Thus the main parts of the various set theories
we consider  differ only with respect to the power of their
comprehension scheme. This, in turn, depends 
only on the safety relation
used by each. 
\item It is easy to see (see \cite{Av04})
that our assumptions concerning the underlying logic 
 and the comprehension schema together imply that the above
formulation of the extensionality axiom is equivalent to the more
usual one:
$\forall z (z\in x \leftrightarrow z\in y)\to x=y$.
\item The first two axioms immediately entail the following two
principles (where $t$ is an arbitrary term):
\bi
\item $\{x\mid x\in t\}=t$ (provided $x\not\in Fv(t)$)
\item $t\in\{x\mid \fe\}\leftrightarrow\fe[x\mapsto t]$ (provided
$t$ is free for $x$ in $\fe$)
\ei
\ms
These principles are counterparts of
the reduction rules $(\eta)$ and $(\beta)$ (respectively) from
the $\lambda$-calculus. Like their counterparts, they are
designed to be used as simplification
rules (at least in the solution of elementary problems).
\ee

\section{The Most Basic System}

Our most basic system is the one which corresponds
to the minimal safety relation (in a language without $TC$).
For the reader convenience, we explicitly present the
definition of this relation:

\begin{definition}
\label{RST-safety}
The relation $\succ_{RST}$ is inductively defined as follows:
\be
\item $\fe\succ_{RST}\emptyset$ if $\fe$ is atomic.
\item $\fe\succ_{RST}\{x\}$ if $\fe\in\{x=t,t=x,x\in t, x\in x\}$, and
 $x\not\in Fv(t)$.
\item $\neg\fe\succ_{RST}\emptyset$ if $\fe\succ_{RST}\emptyset$.
\item $\fe\vee \psi\succ_{RST}X$
if $\fe\succ_{RST}X$ and $\psi\succ_{RST}X$.
\item $\fe\wedge \psi\succ_{RST}X\cup Y$ if
$\fe\succ_{RST}X$, $\psi\succ_{RST}Y$, and 
$Y\cap Fv(\fe)=\emptyset$. 
\item $\exists y \fe\succ_{RST}X-\{y\}$ if $y\in X$ and $\fe\succ_{RST}X$.
\ee
\end{definition}

\noindent
We denote by $RST$ (Rudimentary Set Theory) the set theory induced by 
$\succ_{RST}$ (within the framework described above). 
Note that $RST$  without the $\in-$induction schema
can be shown to be equivalent to Gandy's basic set
theory \cite{Ga74}, and to the system called $BST_0$ in
\cite{Sa97}).

\medskip

The following theorem
about $RST$ can easily be proved:

\begin{theorem}
Given an expression $E$ and a finite set
$X$ of variables, it is decidable in polynomial time
whether $E$ is a valid term of $RST$, whether 
it is a valid formula of $RST$, and if the latter
holds, whether $E\succ_{RST}X$. 
\end{theorem}

\begin{note}
The last theorem is of a crucial importance from
implementability point of view, and it obtains
also for all the extensions of $RST$ discussed (explicitly
or implicitly) below. In order to ensure it, we did
not include in the definition of safety relations the natural
condition that if $\fe\succ X$  and $\psi$ is (logically) equivalent
to $\fe$ (where $Fv(\fe)=Fv(\psi)$) then also $\psi\succ X$.
However, we obviously do have that if 
$\fe\succ_{RST}\{x\}$, and $\vd_{RST}\fe\leftrightarrow\psi$,
then $\vd_{RST}x\in\{x\mid\fe\}\leftrightarrow\psi$,
and so $\vd_{RST}\exists Z\forall x.x\in Z\leftrightarrow\psi$.
Again this is true for any system in our framework.
\end{note}

\subsection{The Power of $RST$} 
\label{RST-power}


\noindent
In the language of $RST$ we can introduce as {\em abbreviations} 
most of the standard notations
for sets used in mathematics. Again, all these
abbreviations should  be used in a purely  static way: no justifying
propositions and proofs are needed.  Here are some examples:

\bs

\bi
\item $\emptyset=_{Df}\{x\mid x\in x\}$.
\item $\{t_1,\ld,t_n\}=_{Df}\{x\mid x=t_1\vee\ld\vee x=t_n\}$
(where $x$  is new).
\item $\tup{t,s}=_{Df}\{\{t\},\{t,s\}\}$. 
\item $\tup{t_1,\ldots,t_n}$ is 
$\emptyset$ if $n=0$, $t_1$ if $n=1$,
$\tup{\tup{t_1,\ldots,t_{n-1}},t_n}$ if $n\geq 2$.
\item $\{x\in t\mid \fe\}=_{Df}\{x\mid x\in t\w \fe\}$, provided
$\fe\succ_{RST}\emptyset$.
 (where $x\not\in Fv(t)$).
\item $\{t\mid x\in s\}=_{Df}\{y\mid\exists x.x\in s\w y=t\}$
  (where $y$  is new, and $x\not\in Fv(s)$).
\item $s\times t=_{Df}\{x\mid \exists a \exists b. a\in s \wedge b\in t 
      \wedge x=\tup{a,b}\}$ (where $x,a$ and $b$ are new).
\item $\{\tup{x_1,\ldots,x_n}\mid\fe\}=_{Df}\{z\mid
\exists x_1\ldots\exists x_n.\fe\w z=\tup{x_1,\ld,x_n}\}$, if
$\fe\succ_{RST}\{x_1,\ldots,x_n\}$ and $z\not\in Fv(\fe)$.
\item $s\cap t=_{Df}\{x\mid x\in s \w x \in t\}$ (where $x$  is new).
\item $s\cup t=_{Df}\{x\mid x\in s \vee x \in t\}$ (where $x$  is new).
\item $s-t=_{Df}\{x\mid x\in s \w x \not\in t\}$ (where $x$  is new).
\item $S(x)=_{Df}x\cup\{x\}$
\item $\bigcup t=_{Df}\{x\mid\exists y.y\in t\w x\in y\}$
(where $x$ and $y$ are new).
\item $\bigcap t=_{Df}\{x\mid \exists y(y\in t\w x\in y)
\w \fo y(y\in t\to x\in y)\}$
(where $x,y$ are new).
\item $\iota x\fe=_{Df}\bigcap\{x\mid\fe\}$ (provided $\fe\succ\{x\}$).
\item $P_1(z)=\iota x. \exists v \exists y(v\in z\w x\in v\w y\in v\w
       z=\tup{x,y})$ 
\item $P_2(z)=\iota y. \exists v \exists x(v\in z\w x\in v\w y\in v\w
       z=\tup{x,y})$ 
\item $\lambda x\in s.t=_{Df} \{\tup{x,t}\mid x\in s\}
\mbox{\quad (where } x\not\in Fv(s))$
\item $f(x)=_{Df} \iota y.\exists z\exists v(z\in f\w
 v\in z\w y\in v\w z=\tup{x,y})$
\item $Dom(f)=_{Df}\{x\mid \exists z\exists v\exists y(z\in f\w
 v\in z\w y\in v\w x \in v\w y=f(x)\}$
\item $Rng(f)=_{Df}\{y\mid \exists z\exists v\exists x(z\in f\w
 v\in z\w y\in v\w x \in v\w y=f(x)\}$
\item $f/s=_{Df}\{\tup{x,f(x)}\mid x\in s\}$ \quad (where $x$ is new).
\ei
\bs

\noindent
{\bf Notes}
\be
\item
It is straightforward to check that in all these
abbreviations the right hand side
is a valid term of $RST$ (provided that the terms/formulas
occurring in it are valid terms/well-formed formulas of $RST$).
We explain $s\times t$ by way of example: since $a$ and $b$ are new,
$a\in s\succ_{RST}\{a\}$, and $b \in t\succ_{RST}\{b\}$. Since
$b\not\in Fv(a\in s)$, this implies that 
$a\in s \wedge b\in t\succ_{RST}\{a,b\}$. Similarly,
$a\in s \wedge b\in t\wedge x=\tup{a,b}\succ_{RST}\{a,b,x\}$.
It follows that $\exists a \exists b. a\in s \wedge b\in t
\wedge x=\tup{a,b}\succ_{RST}\{x\}$. Hence our term for $s\times t$
(which is the most natural one) is a valid term of $RST$.
\item It can easily be seen that according to these
definitions, $\bigcap \emptyset=\emptyset$, and so
$\iota x\fe$ denotes $\emptyset$ if
there is no set which satisfies $\fe$, while it denotes
the intersection of all the sets which satisfy $\fe$ otherwise.
In particular: if there is exactly one set which satisfies $\fe$,
and $\fe\succ\{x\}$,
then $\iota x\fe$ denotes this unique set (this fact has already been used
above).
It follows that if $\fe(y_1\nek y_n,x)$ implicitly defines
(in some theory extending the basic theory of our framework)
a function $f_\fe$ such that for all $y_1\nek y_n$, $f_\fe(y_1\nek y_n)$
is the unique $x$ such that  $\fe(y_1\nek y_n,x)$,
and if $\fe\succ\{x\}$, then there is a term in
the language which explicitly
denotes $f_\fe$; no extension of the language is needed
for that.
\item It is easy to see that the usual reduction rules of the typed 
$\lambda$-calculus
follow from the corresponding reduction rules described
in Section \ref{Axioms}. In particular:
$\vd_{RST}a\in s\to(\lambda x\in s.t)(a)=t\{a/x\}$.
\ee

Exact characterizations of the operations that are explicitly 
definable in $RST$, and of the strength of $RST$, are
given in the following theorems and corollary

\begin{theorem} \
\be
\item If $F$ is an n-ary rudimentary function\footnote{The class of
rudimentary set functions was introduced independently
by Gandy (\cite{Ga74}) and Jensen (\cite{Je72}). See also \cite{De84},
Sect. IV.1.} 
then there exists a formula $\fe$ s. t.:
\be
\item $Fv(\fe)=\{y,x_1\nek x_n\}$
\item $\fe\succ_{RST}\{y\}$
\item $F(x_1\nek x_n)=\{y\mid\fe\}$.
\ee
\item If $\fe$ is a formula such that:
\be
\item $Fv(\fe)=\{y_1\nek y_k,x_1\nek x_n\}$
\item $\fe\succ_{RST}\{y_1\nek y_k\}$
\ee
then there exists
a rudimentary function $F$ such that:
$$F(x_1\nek x_n)=\{\tup{y_1\nek y_k}\mid\fe\}$$
\ee
\end{theorem}

\begin{corollary}
If $Fv(\fe)=\{x_1\nek x_n\}$, and
$\fe\succ_{RST}\emptyset$ then $\fe$ defines
a rudimentary predicate $P$. 
Conversely, if $P$ is  rudimentary then there
is a formula $\fe$ such that $\fe\succ_{RST}\emptyset$ and 
$\fe$ defines $P$.
\end{corollary}


\subsection{Generalized Absoluteness} 

For simplicity of presentation, we assume the 
cumulative universe $V$ of $ZF$, and formulate our definitions accordingly.
It is easy to see that $V$ is a model of $RST$ (with the obvious
interpretations of $RST$'s terms).

\begin{definition}
Let ${\cal M}$ be a transitive model of $RST$. Define the relativization
to ${\cal M}$ of the terms and formulas of $RST$ recursively as follows:
\bi
\item $t_{\cal M}=t$ if $t$ is a variable or a constant.
\item $\{x\mid\fe\}_{\cal M}=\{x\mid x\in {\cal M} \wedge \fe_{\cal M}\}$.
\item $(t=s)_{\cal M}=(t_{\cal M}=s_{\cal M})$\quad
      $(t\in s)_{\cal M}=(t_{\cal M}\in s_{\cal M})$.
\item $(\neg\fe)_{\cal M}=\neg\fe_{\cal M}$\quad
      $(\fe\vee\psi)_{\cal M}=\fe_{\cal M}\vee\psi_{\cal M}$.\quad
      $(\fe\wedge\psi)_{\cal M}=\fe_{\cal M}\wedge\psi_{\cal M}$.
\item $(\exists x\fe)_{\cal M}=\exists x(x\in {\cal M}\wedge \fe_{\cal M})$.
\ei
\end{definition}

\begin{definition} 
Let $T$ be an extension of $RST$ such that $V\models T$.
\be
\item Let $t$ be a term, and let $Fv(t)=\{y_1,\ldots,y_n\}$.
We say that $t$ is $T$-{\em absolute} if the following is true (in $V$)
for every transitive model ${\cal M}$ of $T$:
\[\forall y_1\ldots \forall y_n.
y_1\in{\cal M}\wedge\ldots\wedge y_n\in{\cal M}\to t_{\cal M}=t\]
\item Let $\fe$  be a formula,   and let $Fv(\fe)=
\{y_1,\ldots,y_n, x_1,\ldots,x_k\}$. We say that $\fe$ is $T$-{\em absolute
for $\{x_1,\ldots,x_k\}$} if $\{\tup{x_1,\ld,x_k}\mid\fe\}$ is a set
for all values of the parameters $y_1,\ldots,y_n$,
and the following is true (in $V$) for every transitive model
${\cal M}$ of $T$:
\[\forall y_1\ldots \forall y_n.
y_1\in{\cal M}\wedge\ldots\wedge y_n\in{\cal M}\to
[\fe\leftrightarrow(x_1\in{\cal M}\wedge\ldots\wedge x_k\in{\cal M}
\wedge \fe_{\cal M})]\]
\ee
\end{definition}

\noindent
Thus a term is $T$-absolute if it has the same interpretation in all transitive
models of $T$ which contains the values of its parameters, while
a formula is $T$-absolute for $\{x_1,\ldots,x_k\}$ if it has the same extension
(which should be a set) in all transitive models of $T$ which contains the
values of its other parameters. In particular: $\fe$ is $T$-absolute for
$\emptyset$ iff it is absolute relative to $T$ in the usual sense
of set theory (see e.g. \cite{Ku80}), while $\fe$
is $T$-absolute for $Fv(\fe)$ iff it is domain-independent
in the sense of database theory 
for transitive models of $T$.

\begin{theorem} \
\label{abs-theo}
\be
\item Any valid term $t$ of $RST$ is $RST$-absolute.
\item If $\fe\succ_{RST}X$ then $\fe$ is $RST$-absolute for $X$.
\ee
\end{theorem}


\section{Handling the Axioms of $ZF$ and $ZFC$}

\subsection{Subsets, replacement, and Powerset}

The definability of $\{t,s\}$ and of $\bigcup t$
in the language of $RST$ means that the
axioms of pairing and union are provable in $RST$. We turn now to the question
how to deal with the other comprehension axioms of $ZF$ within
the proposed framework. We start with the comprehension axioms
that remain valid if we limit ourselves to hereditarily finite sets.
It can be shown (\cite{Av08}) that each of them can be captured (in
a modular way)  by adding to
the definition of $\succ_{RST}$ a certain syntactic condition. Here
are those conditions:

\bd
\item[Separation:]
$\fe\succ\emptyset$ for every formula $\fe$.
\item[Replacement:]
 $\exists y \fe\wedge\forall y(\fe\rightarrow \psi)\succ X$
if $\psi\succ X$,  and $X\cap Fv(\fe)=\emptyset$.
\item[Powerset:] $\fo y(y\in x\to\fe)\succ (X-\{y\})\cup\{x\}$ if
$\fe\succ X$, $y\in X$, and $x\not\in Fv(\fe)$. 

Another (and perhaps simpler) method
to handle the powerset axiom is to enrich first the language with
a new binary relation $\subseteq$. 
Then add to the definition
of the safety relation the condition: $x\subseteq t\succ_{}\{x\}$ if
$x\not\in Fv(t)$. Finally, 
 add the usual definition of $\subseteq$ in terms of $\in$ as
an extra {\em axiom}: 
$\fo x\fo y(x\subseteq y\lw\fo z(z\in x\to z\in y))$.
Alternatively, 
since $\subseteq$ is now taken as primitive, it might be
more natural to use it as such in our axioms. This means that 
instead of adding the above axiom, it might be preferable to replace 
the single extensionality axiom of $BZF$ with the following three:
%
(Ex1) $x\subseteq y\w y\subseteq x\to x=y$,
(Ex2) $z\in x\w x\subseteq y\to z\in y$, and
(Ex3) $x\subseteq y\vee\exists z(z\in x\w z\not\in y)$.
\ed

\begin{note}
If any of the conditions introduced in this subsection
is used then the counterpart of Theorem \ref{abs-theo}
is not valid for the resulting system. 
Hence these conditions are not coherent
with our initial intuitions
(Thus from the perspective of our framework,
the condition that corresponds to the separation schema 
means that from the point of view of  $ZF$, every formula
defines a ``decidable" relation on the universe $V$ of sets).
As a compensation, we  have the following remarkable property
of the condition that corresponds to replacement (see \cite{Av07}):
\end{note}

\begin{theorem}
Let ${\cal T}$ be a set theory in our framework such that 
the corresponding safety relation $\succ_{{\cal T}}$ satisfies
the condition that corresponds to replacement. Then
for any formula $\fe$ of ${\cal T}$ 
such that $Fv(\fe)=\{ y_1,\ld,y_n,x\})$,
there exists a term $t_\fe$ of ${\cal T}$ such that 
$Fv(t_\fe)=\{ y_1,\ld,y_n\}$,
and
$$\vd_{{\cal T}}\fo y_1,\ld,y_n\exists ! x\fe\to\fo y_1,\ld,y_n(\fe[x\mapsto
t_\fe])$$

\end{theorem}

\subsection{The Axiom of Infinity}

Next we turn to the axiom of Infinity --- the only comprehension axiom that
necessarily takes us out of the realm of finite sets. As long as we
stick to first-order languages, it seems
impossible to incorporate it into our systems by
just imposing new simple syntactic conditions
on the safety relation.
Instead, the best way to capture it
is to add to the basic signature a new
constant $HF$ (interpreted as the collection ${\cal HF}$ of hereditarily
finite sets)
together with the following counterparts of {\em Peano's axioms:}
\be
\item $\emptyset\in HF$
\item $\fo x\fo y. x\in HF\w y\in HF\to x\cup\{y\}\in HF$
\item $\fe(0)\w(\fo x\fo y.\fe(x)\w\fe(y)\to\fe(x\cup\{y\})\to
\fo x\in HF.\fe(x)$
\ee

\begin{definition}
{\rm $RST\omega$ is the theory which is obtained from $RST$
by the addition of the constant $HF$  and the above 
counterparts of  Peano's axioms.
}
\end{definition}

On the other hand, if a language with $TC$ is used, 
then we get the infinity axiom for free, since
both ${\cal HF}$  and the set $\omega$ of the finite ordinals 
are definable
in this extended language by valid terms (see \cite{Av10}). Thus 
the one that defines $\omega$ is 
$\omega=\{y\mid\exists x. x=\emptyset\w
(TC_{x,y}y=\{z\mid z=x\vee z\in x\})(x,y)\}$.

\begin{definition}
{\rm Let $\succ_{PZF}$ be the minimal safety relation in a language
with $TC$ (note that the only difference between $\succ_{PZF}$
and $\succ_{RST}$ is the extra clause for $TC$). We denote by $PZF$
(predicative set theory) the set theory induced by $\succ_{PZF}$
within our framework.}
\end{definition}

\begin{note}
An important property of $RST\omega$ and $PZF$ is that
Theorem \ref{abs-theo} does remain valid if instead of $RST$ 
we consider either of them. Hence these systems {\em are}
coherent with our initial motivations and intuitions.
\end{note} 

\subsection{The Axiom of Choice}
The full set theory ZFC has one more axiom, which does not fit into
the formal framework described above: $AC$ (the axiom of  choice).
It seems that the most natural
way to incorporate it into our framework is by further extending the
set of terms, using Hilbert's $\varepsilon$ symbol,
together with its usual characterizing axiom (which is equivalent
to the axiom of global choice):
$\exists x\fe\to\fe[x\mapsto \varepsilon x\fe]$.
It should be noted that this move is not in line with our
stated goal of employing only standard notations used in textbooks,
but some price should be paid for including the axiom of choice
in a system.

\section{Structures and Computations}

Let ${\cal T}$ be a theory  formulated within the classical
part of our framework. From
the Platonist point of view  its set
of closed terms 
induces some subset ${\cal S}({\cal T})$
of the universe $V$ of sets. The identity of ${\cal S}({\cal T})$
depends only on the {\em language} of ${\cal T}$ and on the interpretations
of the symbols its signature has in addition to
$\in,=$, and $\subseteq$ (if such symbols exist). It does not
depend on its axioms. In addition, for any transitive model
${\cal M}$ of ${\cal T}$, ${\cal S}({\cal T})$ determines some subset
${\cal M}({\cal T})$ of ${\cal M}$ (which might not be an element of
${\cal M}$). Now a theory ${\cal T}$ is computationally
interesting if the set ${\cal S}({\cal T})$ it induces
is a ``universe" in the sense that it is a transitive
model of ${\cal T}$. According to our guiding ideas, such a theory
${\cal T}$ and its model  ${\cal S}({\cal T})$ have
a special significance from a computational point
of view if the identity of the latter 
is {\em absolute} in the sense
that ${\cal M}({\cal T})={\cal S}({\cal T})$ for any transitive model
${\cal M}$ of ${\cal T}$ (implying that ${\cal S}({\cal T})$
is actually a  {\em minimal} transitive model of ${\cal T}$).
From results in \cite{Av10} it follows that
at least the following theories 
have both properties:

\bd
\item[$RST$:] 
Its minimal model 
${\cal S}(RST)$ is identical to ${\cal HF}$, which is
$J_1$ in Jensen's hierarchy (\cite{Je72,De84}), and $L_\omega$ in
G\"odel hierarchy (\cite{Go40,De84}) of constructible sets.
\item[$RST\omega$:] 
Its minimal model ${\cal S}(RST\omega)$
is $J_2$ in Jensen's hierarchy.
\item[$PZF_{{\cal TCL}}$:] 
Its minimal model is $J_{\omega^\omega}=L_{\omega^\omega}$.
\ed

\nocite{*}
\bibliographystyle{eptcs}
\bibliography{set-theories}
 
\end{document}